\documentclass{article}
\usepackage{spconf}
\usepackage{amsmath}
\usepackage{graphicx}
\usepackage{array}
\usepackage{url}
\usepackage{multirow}
\usepackage{balance}
\usepackage{bbm}
\usepackage{cite}
\usepackage{xcolor}
\usepackage{caption}
\usepackage{subfigure}
\usepackage{setspace}
\usepackage[T1]{fontenc}

\title{Speaker recognition with two-step multi-modal deep cleansing}

\name{Ruijie Tao$^{1}$, Kong Aik Lee$^{2}$, Zhan Shi$^{3}$ and Haizhou Li$^{1,3}$ }

\address{
  $^{1}$College of Design and Engineering, National University of Singapore, Singapore\\
  $^{2}$Institute for Infocomm Research, A$^\star$STAR, Singapore~~~~~~\\
  $^{3}$The Chinese University of Hong Kong, Shenzhen, China}
 
\begin{document}
\topmargin=0mm
\maketitle

\begin{abstract}
Neural network-based speaker recognition has achieved significant improvement in recent years. A robust speaker representation learns meaningful knowledge from both hard and easy samples in the training set to achieve good performance. However, noisy samples (i.e., with wrong labels) in the training set induce confusion and cause the network to learn the incorrect representation. In this paper, we propose a two-step audio-visual deep cleansing framework to eliminate the effect of noisy labels in speaker representation learning. This framework contains a coarse-grained cleansing step to search for the peculiar samples, followed by a fine-grained cleansing step to filter out the noisy labels. Our study starts from an efficient audio-visual speaker recognition system, which achieves a close to perfect equal-error-rate (EER) of 0.01\%, 0.07\% and 0.13\% on the Vox-O, E and H test sets. With the proposed multi-modal cleansing mechanism, four different speaker recognition networks achieve an average improvement of 5.9\%. Code has been made available at: \textcolor{magenta}{\url{https://github.com/TaoRuijie/AVCleanse}}.
\end{abstract}
\vspace{-3mm}
 
\section{Introduction}
\label{sec:1}
 
\begin{figure*}[!ht]
    \centering
    \includegraphics[width=\linewidth]{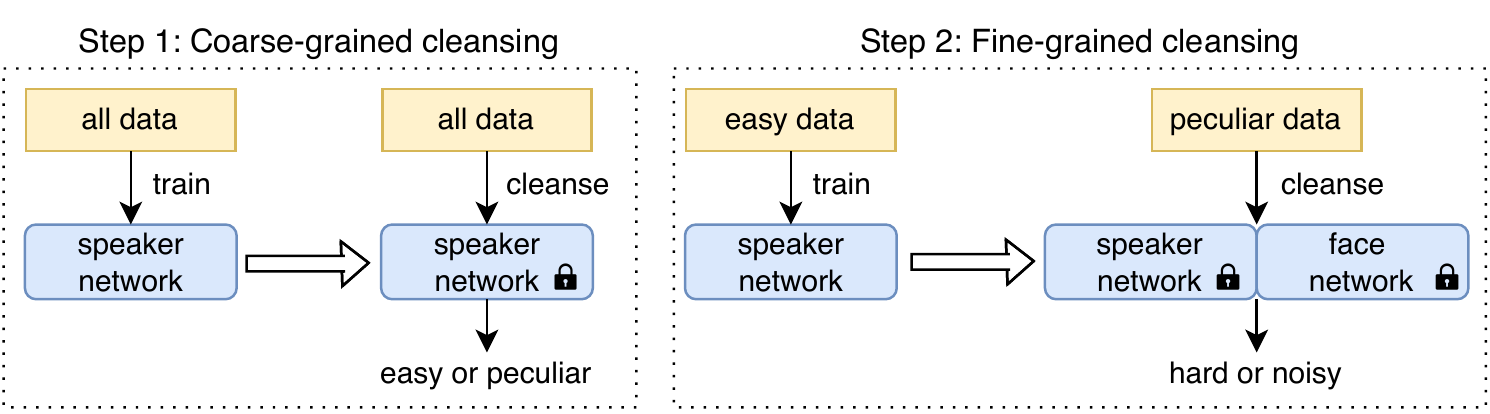}
    \caption{The proposed two-step audio-visual deep cleansing framework. The lock represents that the network is frozen.}
    \label{fig:framework}
\end{figure*}

Automatic speaker recognition aims to distinguish a legitimate user from imposters based on their voices~\cite{lee2011joint, APSIPA_realism}. Over the last decade, deep learning-based speaker representation, such as x-vector~\cite{xvectors}, xi-vector~\cite{lee2021xi}, and network architectures, such as ResNet~\cite{he2016deep} and ECAPA-TDNN~\cite{desplanques2020ecapa}, have achieved remarkable performance by training on the large-scale speech dataset. To further enhance the performance, previous works usually focus on the network structure~\cite{Tianchi_icassp22}, loss function~\cite{chung2020defence} and back-end score calibration~\cite{Thienpondt2021IntegratingFT}. However, the existence of noisy labels in the training set has not been taken care of.

Large-scale dataset for speaker recognition typically consists of thousands of speakers and over millions of samples, whereby utterances from the same person share the same speaker labels. These datasets are typically collected from the Internet within automatic pipeline~\cite{Voxceleb, Voxceleb2}. However, it is not surprising to find that utterances assigned with the same label might actually come from different speakers, i.e., the noisy labels problem. From the study of noisy-label learning, training data can be divided into three categories according to the learning difficulty~\cite{castells2020superloss, huang2019o2u}: easy, hard and noisy samples. Note that we use the term noisy samples referring to samples with wrong labels following the terminology in ~\cite{Voxceleb}. Due to the memorization effects~\cite{arpit2017closer}, neural networks tend to fit the easy samples and converge rapidly in the early stage of training. In the later stage of training, the network has to distil correct but difficult knowledge from the peculiar samples. However, these peculiar samples contain hard and noisy samples, which deters the learning process.

Prior works dealing with noisy labels in speaker recognition include: training the speaker network and manually setting the threshold to filter out the unusual utterances~\cite{qin2022simple}; dropping out the samples with large training loss as the noisy data in self-supervised learning~\cite{tao2021self}. However, there are two major problems with the existing approach. In terms of quality, it is challenging for the speaker network to recognize the difficult utterances~\cite{li2022pay}. In terms of logic, the speaker network has been trained on the entire dataset (including noisy ones) and push them to their class centre. Using the same network to decide the noisy ones is not a good choice.

In this paper, we propose a two-step multi-modal deep cleansing framework to solve the above mentioned problems. Firstly, we do a coarse-grained cleansing based on the speech modality only, which divides the training data into easy and peculiar samples. Secondly, we train a new speaker recognition network based on easy samples to do a fine-grained cleansing, which divides the peculiar samples into hard and noisy ones. Since the network does not train on the noisy samples, using it to do the cleansing is more reliable. On the other hand, from the biometric recognition study~\cite{qian2021audio, HLT_SRE2019}, face image and speech utterance can provide complementary identity information. Motivated by this, we use the face recognition network in the second step to boost the system. 

Our contribution can be summarized as follows. Firstly, we propose a robust audio-visual speaker recognition system which can achieve close-to-perfect verification. Secondly, a two-step audio-visual cleansing framework is designed to filter out the noisy data in the training set. Thirdly, four speaker recognition networks are trained on the original and cleansed dataset to show the impact of our method.

\section{Audio-visual speaker recognition}
\label{sec:2}
In this section, we propose an efficient audio-visual speaker recognition system containing both speaker and face modalities since a reliable identity recognition system is the foundation of noisy sample cleansing.

For training, the speaker network is used to extract the speaker embedding from the input utterance. This embedding contains the characteristics of the speaker's voice. Then it follows a speaker classifier to distinguish the utterances from different speakers with an AAM-softmax loss~\cite{deng2019arcface}. Similarly, the face network takes one face image as the input and outputs the face embedding. This embedding is trained with a face classifier~\cite{schroff2015facenet}. For testing, speaker and face embeddings are used together to enhance the verification performance.

Compared with previous works~\cite{shon2019noise, sari2021multi}, we take two strategies to simplify the system and improve the stability. Firstly, we align all the faces with the detected landmarks during preprocessing~\cite{zhang2016joint} since the unaligned faces in the training set make recognition harder~\cite{kowalski2017deep}. Secondly, existing approaches usually attempt to combine the speaker and face modality by the attention mechanism~\cite{qian2021audio, chen2020multi}. However, we argue that training two modalities separately and directly concatenating two embeddings for testing is an effective and convenient solution.~\cite{sari2021multi, HLT_SRE2019}

\section{Audio-visual deep cleansing}
\label{sec:3}
For audio-visual deep cleansing, we consider a training set with $N$ video clips. Each video clip consists of synchronized speech utterance and face frames. And, there exists a few video clips with wrong labels, i.e., noisy samples. Our proposed audio-visual two-step deep cleansing framework is shown in Fig~\ref{fig:framework} to discover these noisy samples. 

\subsection{Step 1: Coarse-grained cleansing}
\label{sec:3.1}
Firstly, we design coarse-grained cleansing to divide the training set into easy and peculiar samples to narrow down the selection of noisy data. The main concern of this step is successfully dividing all the noisy samples into the peculiar class. It is noted that this step uses speech modality only since face modality can only assist in deciding the correctness of samples.

Firstly, we train a speaker network with all the utterances in the training set. Then we fix this network's parameters and extract the speaker embeddings from the entire training set. These embeddings are represented as $s_1, \cdots, s_N$. The corresponding speaker labels are annotated as $c_1, \cdots, c_N, c_i \in \{1,2, \cdots, K\}$, where $K$ is the number of speakers in the training set. Then we compute the average cosine similarity between the speaker embedding of each sample and the other samples from the same speaker. The score $x_i$ of the embedding $s_i$ can be represented as:
\begin{equation}
    \label{e1}
    x_i = \frac{1}{M_k}\sum_{j=1}^{N}\mathbbm{1}_{\substack{c_i = c_j}}{cos(s_i, s_j)} 
\end{equation}
Here $M_k$ is the number of samples in the $k^{th}$ class. $\mathbbm{1}$ is an indicator function evaluating 1 when $c_i = c_j$. For the computed scores, we set a threshold $\tau$. Samples with scores smaller than $\tau$ are the peculiar samples; otherwise, they are deemed to be easy samples.

\subsection{Step 2: Fine-grained cleansing}
\label{sec:3.2}
As mentioned earlier, the network without training on the noisy samples can provide an objective and accurate representation. Motivated by that, we train the network without noisy samples in fine-grained cleansing to filter out noisy samples.

\subsubsection{Speaker and face network}
Firstly, we train a new speaker network with the easy samples found from the coarse-grained cleansing. Then considering sufficient annotated face datasets, the pre-trained face recognition network is applied to our multi-modal cleansing proposal. It is noted that training a face network with the images of the found easy samples is also a reasonable solution.

\subsubsection{Decision boundary}
Then we train a classifier in the two-dimensional score space to separate noisy samples from the hard samples. The binary classifier is trained on a validation set consisting of target trials~\cite{Voxceleb, Voxceleb2}, i.e., two video clips from the same speakers, and the imposter trials, i.e., two video clips are from different speakers. We can compute the speaker and face cosine similarity for each trial using our multi-modal system. Then an SVM~\cite{cortes1995support} is learnt on these two-dimensional scores and using the ground-truth labels in the validation. The SVM decision boundary can efficiently distinguish trials into the target and the imposter ones. Target and imposter trials are associated with the clean and noisy samples, respectively.

\subsubsection{Deep cleansing}
In this stage, we freeze the networks to extract the speaker and face embeddings of all the samples in the training set. Speaker and face embeddings are represented as $s_1, \cdots, s_N$ and $f_1, \cdots, f_N$, respectively. We compute the average cosine similarity between the embedding of each sample and the other samples from the same class. The speaker scores $x_i$ of the embedding $s_i$ is similar to that in (\ref{e1}). The face score $y_i$ of the embedding $f_i$ is computed as:
\begin{equation}
    \label{e2}
    y_i = \frac{1}{M_k}\sum_{j=1}^{N}\mathbbm{1}_{\substack{c_i = c_j}}{cos(f_i, f_j)} 
\end{equation}

Finally, we apply the learnt SVM to predict the correctness of each training sample using $x_i$ and $y_i$. The samples predicted as the target trials are defined as clean data; otherwise, they are noisy data.

\section{Experimental setup}
\label{sec:4}
Our training set is the VoxCeleb2~\cite{Voxceleb2}, an audio-visual dataset derived from YouTube interviews. It contains 1,091,724 video clips from 5,994 speakers, and each video clip has one synchronized and visible talking face.

For audio-visual speaker recognition, the original VoxCeleb2 is used for training. Here we also report the performance of the face network trained on the large face dataset Glint360K~\cite{an2021partial} to show the effectiveness of face modality. The speaker network is the ECAPA-TDNN with a large channel size equal to 1024 (refer to ECAPA-L)~\cite{desplanques2020ecapa}. The face network is the ResNet18 (training on VoxCeleb2) and ResNet50 (training on Glint360K)~\cite{deng2019arcface, liang2018learning}. We select Vox1-O set for validation, Vox1-E and Vox1-H for testing~\cite{Voxceleb}.

\begin{table}[!htb]
\setlength{\tabcolsep}{4.5pt}
\caption{The EER (\%) of audio-visual speaker recognition. `-Vox2' denotes training on VoxCeleb2 dataset, '-Glint' denotes training on Glint360K dataset.}
\label{tab1}
\begin{spacing}{1}
\begin{tabular}{ccccc}
\hline
\textbf{Modality}                & \textbf{System}                            & \textbf{Vox1-O} & \textbf{Vox1-E} & \textbf{Vox1-H} \\ \hline
\multirow{5}{*}{\textbf{Speech}}  & Sari et al.~\cite{sari2021multi} & 2.20 & - & -        \\ 
                                  & Qian et al.~\cite{qian2021audio} & 1.62 & 1.75 & 3.16        \\ 
                                  & Chen et al.~\cite{chen2020multi} & 2.31 & 2.23 & 3.78        \\ 
                                  & \textbf{(1) Ours-Vox2}           & \textbf{1.02} & \textbf{1.23} & \textbf{2.36}        \\ \hline
\multirow{6}{*}{\textbf{Face}}    & Sari et al.~\cite{sari2021multi} & 3.90 & - & -        \\ 
                                  & Qian et al.~\cite{qian2021audio} & 3.04 & 2.18 & 4.23        \\
                                  & Chen et al.~\cite{chen2020multi} & 2.26 & 1.54 & 2.37        \\  
                                  & \textbf{(2) Ours-Vox2}           & \textbf{0.97} & \textbf{0.81} & \textbf{1.16}        \\ 
                                  & \textbf{(3) Ours-Glint}          & \textbf{0.03} & \textbf{0.07} & \textbf{0.09}        \\ \hline
\multirow{5}{*}{\textbf{Fusion}}  & Sari et al.~\cite{sari2021multi} & 0.90 & - & -        \\ 
                                  & Qian et.al.~\cite{qian2021audio} & 0.71 & 0.48 & 0.85       \\
                                  & Chen et al.~\cite{chen2020multi} & 0.59 & 0.43 & 0.74        \\
                                  & \textbf{(1) + (2)}               & \textbf{0.16} & \textbf{0.23} & \textbf{0.42}        \\
                                  & \textbf{(1) + (3)}               & \textbf{0.01}   & \textbf{0.07}   & \textbf{0.13} \\ \hline
\end{tabular}
\end{spacing}
\end{table}

For audio-visual deep cleansing, the speaker network is an ECAPA-L, and the face network is a ResNet50 network. In fine-grained cleansing, we set the threshold $\tau$ to divide 92\% of training data as the easy samples. Here we only need to ensure that no noisy samples are involved~\cite{qin2022simple}. In coarse-grained cleansing, the face network is trained on Glint360K~\cite{an2021partial}. Here only the cleansed samples are used to decide the class centre, so we repeat five rounds to compute each sample's similarity score and find clean samples.

To show the impact of our audio-visual deep cleansing framework, we train four speaker networks with and without cleansing and compare their performances. The networks include x-vector~\cite{xvectors}, ResNet34~\cite{he2016deep}, ECAPA-TDNN~\cite{desplanques2020ecapa} with a small channel size equal to 512 (refer to ECAPA-S) and the ECAPA-L. All the experiments are repeated three times with the same setting. Vox1-O, E and H are used for in-domain evaluation and CnCeleb-VoxSRC22~\cite{fan2020cn}\footnote{\url{https://www.robots.ox.ac.uk/~vgg/data/voxceleb/data_workshop_2022/Track3_validation_trials.txt}} is used for cross-domain evaluation.

During training, we apply data augmentation for speaker and face networks to boost performance~\cite{MUSAN, RIRS}. During the evaluation, all test sets provide a set number of trials, each containing two samples. For single modality, the cosine similarity between the speaker embedding (from the entire utterance) or the face embedding (from five face frames) of the given trial is calculated. For multi-modal, the combined speaker and face embedding is applied. The performance metric is the equal error rate (EER).

\section{Results and Analysis}
\label{sec:5}

\subsection{Audio-visual speaker recognition}
\label{sec:5.1}

\begin{table}[!htb]
\setlength{\tabcolsep}{5.6pt}
\caption{In-domain evaluation EER(\%) of the speaker networks trained on the original and cleansed VoxCeleb2.}
\label{tab2}
\begin{spacing}{1}
\begin{tabular}{ccccc}
\hline
\textbf{Network} & \textbf{Method} & \textbf{Vox1-O} & \textbf{Vox1-E} & \textbf{Vox1-H} \\ \hline
\multirow{3}{*}{\textbf{X-vector}} & w/o cleanse  & 2.20  & 2.32  & 4.06  \\
                                   & with cleanse & 2.09  & 2.13  & 3.76  \\
                                   & \textbf{$\Delta$}   & \textbf{5.0\%} & \textbf{8.2\%} & \textbf{7.4\%} \\ \hline
\multirow{3}{*}{\textbf{ResNet34}} & w/o cleanse  & 1.31  & 1.41  & 2.58  \\
                                   & with cleanse & 1.24  & 1.28  & 2.52  \\
                                   &\textbf{ $\Delta$}   & \textbf{5.3\%} & \textbf{9.2\%} & \textbf{2.3\%} \\ \hline
\multirow{3}{*}{\textbf{ECAPA-S}}  & w/o cleanse  & 1.24  & 1.34  & 2.49  \\
                                   & with cleanse & 1.17  & 1.28  & 2.37  \\
                                   & \textbf{$\Delta$}   & \textbf{5.6\%} & \textbf{4.5\%} & \textbf{4.8\%} \\ \hline
\multirow{3}{*}{\textbf{ECAPA-L}}  & w/o cleanse  & 1.02  & 1.23  & 2.36  \\
                                   & with cleanse & 0.93  & 1.18  & 2.22  \\
                                   & \textbf{$\Delta$}   & \textbf{8.8\%} & \textbf{4.1\%} & \textbf{5.9\%} \\ \hline       
\end{tabular}
\end{spacing}
\end{table}
\vspace{-2mm}
First, we report the performance of our audio-visual speaker recognition system in Tab~\ref{tab1}. Both speaker and face networks perform better than existing approaches when trained on the VoxCeleb2 dataset. Here the face recognition network can achieve even 0.03\% EER when trained on Glint360K. For multi-modal verification, we obtain 0.16\% and 0.01\% EER on Vox1-O when the face network is trained on VoxCeleb2 and Glint360K, respectively. So our audio-visual system can accurately define the person's identity to guarantee the cleansing process.

\subsection{Audio-visual deep cleansing}
\label{sec:5.2}
Then we compare the performance of the speaker network trained on the original VoxCeleb2 and the VoxCeleb2 with our deep cleansing. In Tab~\ref{tab2}, for in-domain evaluation on VoxCeleb1, our audio-visual deep cleansing can remove the side-effect of the noisy data and boost the speaker recognition system with an average improvement of 5.9\%. In Tab~\ref{tab3}, for cross-domain evaluation on CnCeleb-VoxSRC22, our approach achieve an average improvement of 3.2\%, which proves the network obtains a stronger generality.

\begin{table}[!h]
\setlength{\tabcolsep}{3.1pt}
\caption{Cross-domain evaluation EER(\%) of the speaker networks trained on the original and cleansed VoxCeleb2.}
\label{tab3}
\begin{spacing}{1}
\begin{tabular}{ccccc}
\hline
\textbf{Method} & \textbf{X-vector} & \textbf{ResNet34} & \textbf{ECAPA-S} & \textbf{ECAPA-L} \\ \hline
\textbf{w/o cleanse}  & 17.00 & 14.90 & 18.15 & 19.86 \\ 
\textbf{with cleanse} & 16.34 & 14.64 & 17.39 & 19.26 \\ 
$\mathbf{\Delta}$   & \textbf{3.9\%} & \textbf{1.7\%} & \textbf{4.2\%} & \textbf{3.0\%} \\ \hline
\end{tabular}
\end{spacing}
\end{table}
\vspace{-2mm}

\subsection{Visualization of results}
\label{sec:5.3}

\begin{figure}[!ht]
    \centering
    \includegraphics[width=\linewidth]{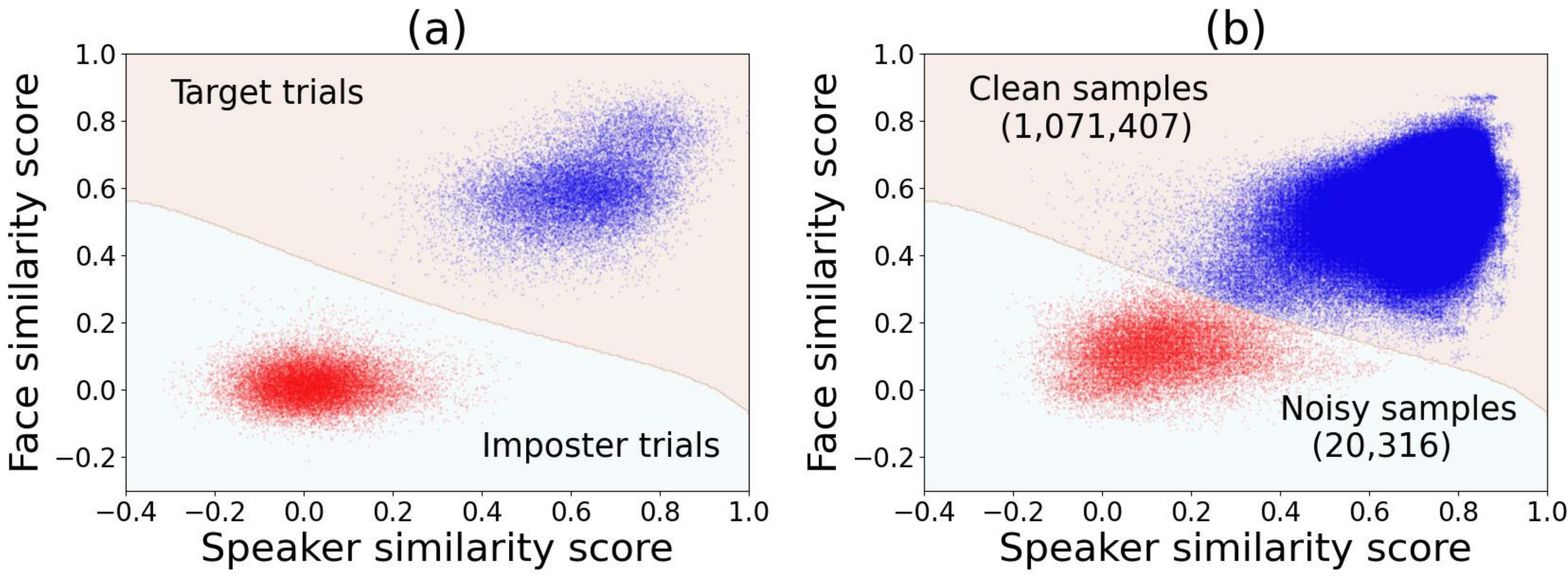}
    \caption{Visualization of (a) audio-visual speaker recognition on Vox1-O; (b) audio-visual deep cleansing on VoxCeleb2.}
    \label{fig:res}
\end{figure}
\vspace{-2mm}
We visualize our results in Fig~\ref{fig:res}. The left panel (a) represents audio-visual speaker recognition on Vox1-O. Each point denotes a test trial. The X-axis and Y-axis represent the speaker and face similarity score between the two samples in each trial, respectively. Target trials have high audio-visual scores, and imposter trials are the opposite, so the decision boundary is clear and reliable. The right panel (b) represents audio-visual deep cleansing on VoxCeleb2 with the same boundary. Each point represents a training sample. The similarity score is between this point and the cleansed samples from the same speaker. Here most of the noisy samples have a very different representation to the mainstream samples in their class since they have a low audio-visual score. Our method finds 1.9\% noisy samples on VoxCeleb2.

\label{sec:5.4}
\begin{figure}[!ht]
    \centering
    \includegraphics[width=\linewidth]{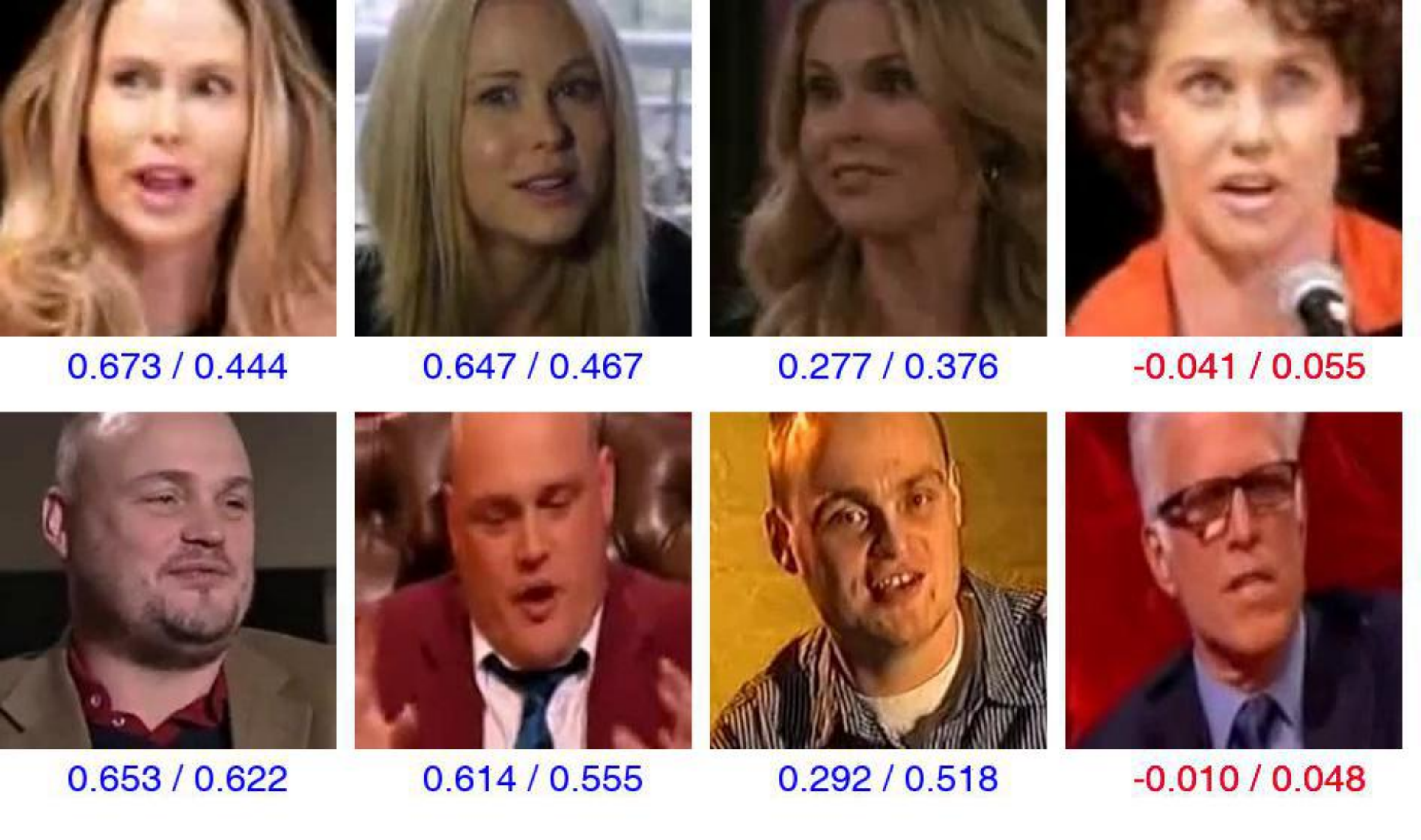}
    \caption{Visualization of the clean and noisy samples with the speaker and face similarity score.}
    \label{fig:vis}
\end{figure}
\vspace{-2mm}
In Fig~\ref{fig:vis}, each row contains four face images from the samples with the same speaker label, the left three are from the clean samples, and the right one is from the noisy sample. Both speaker and face similarity score has been marked below. Visual modality can assist speech modality in finding noisy samples.

\vspace{-2mm}
\section{Conclusion}
\label{sec:6}
In this paper, we design an audio-visual speaker recognition system that achieves close-to-perfect verification on the VoxCeleb1 test sets. A two-step audio-visual deep cleansing framework is proposed to automatically pick up noisy training sounds and strengthen the speaker recognition network. We observe that noisy samples (i.e., with wrong labels) are commonplace in large-scale datasets, and 5.9\% of performance improvement could be achieved by simply removing a considerable percentage of noisy samples from the training set. In future work, we will study an end-to-end approach to combine the training and cleansing steps.

\clearpage
\balance
\footnotesize
\bibliographystyle{IEEEbib}
\bibliography{refs}

\begin{thebibliography}{10}

\bibitem{lee2011joint}
Kong~Aik Lee, Anthony Larcher, Helen Thai, Bin Ma, and Haizhou Li,
\newblock ``Joint application of speech and speaker recognition for automation
  and security in smart home,''
\newblock in {\em Interspeech}, 2011, pp. 3317--3318.

\bibitem{APSIPA_realism}
Rohan~Kumar Das and S.~R.~Mahadeva Prasanna,
\newblock ``Investigating text-independent speaker verification from
  practically realizable system perspective,''
\newblock in {\em Asia-Pacific Signal and Information Processing Association
  Annual Summit and Conference (APSIPA ASC)}, 2018, pp. 1483--1487.

\bibitem{xvectors}
D.~Snyder, D.~Garcia-Romero, G.~Sell, D.~Povey, and S.~Khudanpur,
\newblock ``X-vectors: robust {DNN} embeddings for speaker recognition,''
\newblock in {\em 2018 IEEE International Conference on Acoustics, Speech and
  Signal Processing (ICASSP)}, 2018, pp. 5329--5333.

\bibitem{lee2021xi}
Kong~Aik Lee, Qiongqiong Wang, and Takafumi Koshinaka,
\newblock ``Xi-vector embedding for speaker recognition,''
\newblock {\em IEEE Signal Processing Letters}, vol. 28, pp. 1385--1389, 2021.

\bibitem{he2016deep}
Kaiming He, Xiangyu Zhang, Shaoqing Ren, and Jian Sun,
\newblock ``Deep residual learning for image recognition,''
\newblock in {\em Proceedings of the IEEE conference on computer vision and
  pattern recognition}, 2016, pp. 770--778.

\bibitem{desplanques2020ecapa}
Brecht Desplanques, Jenthe Thienpondt, and Kris Demuynck,
\newblock ``{ECAPA-TDNN: Emphasized Channel Attention, propagation and
  aggregation in TDNN based speaker verification},''
\newblock in {\em Interspeech}, 2020, pp. 3830--3834.

\bibitem{Tianchi_icassp22}
Tianchi Liu, Rohan~Kumar Das, Kong~Aik Lee, and Haizhou Li,
\newblock ``{MFA: TDNN} with multi-scale frequency-channel attention for
  text-independent speaker verification with short utterances,''
\newblock in {\em IEEE International Conference on Acoustics, Speech and Signal
  Processing (ICASSP)}, 2022, pp. 7517--7521.

\bibitem{chung2020defence}
Joon~Son Chung, Jaesung Huh, Seongkyu Mun, Minjae Lee, Hee~Soo Heo, Soyeon
  Choe, Chiheon Ham, Sunghwan Jung, Bong-Jin Lee, and Icksang Han,
\newblock ``In defence of metric learning for speaker recognition,''
\newblock in {\em Interspeech}, 2020, pp. 2977--2981.

\bibitem{Thienpondt2021IntegratingFT}
Jenthe Thienpondt, Brecht Desplanques, and Kris Demuynck,
\newblock ``Integrating frequency translational invariance in {TDNNs} and
  frequency positional information in {2D} {ResNets} to enhance speaker
  verification,''
\newblock in {\em Interspeech}, 2021.

\bibitem{Voxceleb}
Arsha Nagrani, Joon~Son Chung, and Andrew Zisserman,
\newblock ``{VoxCeleb}: A large-scale speaker identification dataset,''
\newblock in {\em Interspeech}, 2017, pp. 2616--2620.

\bibitem{Voxceleb2}
Joon~Son Chung, Arsha Nagrani, and Andrew Zisserman,
\newblock ``{VoxCeleb2}: Deep speaker recognition,''
\newblock in {\em Interspeech}, 2018, pp. 1086--1090.

\bibitem{castells2020superloss}
Thibault Castells, Philippe Weinzaepfel, and Jerome Revaud,
\newblock ``Superloss: A generic loss for robust curriculum learning,''
\newblock {\em Advances in Neural Information Processing Systems}, vol. 33, pp.
  4308--4319, 2020.

\bibitem{huang2019o2u}
Jinchi Huang, Lie Qu, Rongfei Jia, and Binqiang Zhao,
\newblock ``O2u-net: A simple noisy label detection approach for deep neural
  networks,''
\newblock in {\em Proceedings of the IEEE/CVF international conference on
  computer vision}, 2019, pp. 3326--3334.

\bibitem{arpit2017closer}
Devansh Arpit, Stanis{\l}aw Jastrz{\k{e}}bski, Nicolas Ballas, David Krueger,
  Emmanuel Bengio, Maxinder~S Kanwal, Tegan Maharaj, Asja Fischer, Aaron
  Courville, Yoshua Bengio, et~al.,
\newblock ``A closer look at memorization in deep networks,''
\newblock in {\em International conference on machine learning}. PMLR, 2017,
  pp. 233--242.

\bibitem{qin2022simple}
Xiaoyi Qin, Na~Li, Chao Weng, Dan Su, and Ming Li,
\newblock ``Simple attention module based speaker verification with iterative
  noisy label detection,''
\newblock in {\em IEEE International Conference on Acoustics, Speech and Signal
  Processing (ICASSP)}, 2022, pp. 6722--6726.

\bibitem{tao2021self}
Ruijie Tao, Kong~Aik Lee, Rohan~Kumar Das, Ville Hautam{\"a}ki, and Haizhou Li,
\newblock ``Self-supervised speaker recognition with loss-gated learning,''
\newblock in {\em IEEE International Conference on Acoustics, Speech and Signal
  Processing (ICASSP)}, 2022, pp. 6142--6146.

\bibitem{li2022pay}
Lantian Li, Di~Wang, and Dong Wang,
\newblock ``Pay attention to hard trials,''
\newblock {\em arXiv preprint arXiv:2209.04687}, 2022.

\bibitem{qian2021audio}
Xinyuan Qian, Alessio Brutti, Oswald Lanz, Maurizio Omologo, and Andrea
  Cavallaro,
\newblock ``Audio-visual tracking of concurrent speakers,''
\newblock {\em IEEE Transactions on Multimedia}, 2021.

\bibitem{HLT_SRE2019}
Rohan~Kumar Das, Ruijie Tao, Jichen Yang, Wei Rao, Cheng Yu, and Haizhou Li,
\newblock ``Hlt-nus submission for 2019 nist multimedia speaker recognition
  evaluation,''
\newblock in {\em 2020 Asia-Pacific Signal and Information Processing
  Association Annual Summit and Conference (APSIPA ASC)}, 2020, pp. 605--609.

\bibitem{deng2019arcface}
Jiankang Deng, Jia Guo, Niannan Xue, and Stefanos Zafeiriou,
\newblock ``Arc{F}ace: Additive angular margin loss for deep face
  recognition,''
\newblock in {\em IEEE/CVF Conference on Computer Vision and Pattern
  Recognition (CVPR)}, 2019, pp. 4690--4699.

\bibitem{schroff2015facenet}
Florian Schroff, Dmitry Kalenichenko, and James Philbin,
\newblock ``Facenet: A unified embedding for face recognition and clustering,''
\newblock in {\em Proceedings of the IEEE conference on computer vision and
  pattern recognition}, 2015, pp. 815--823.

\bibitem{shon2019noise}
Suwon Shon, Tae-Hyun Oh, and James Glass,
\newblock ``Noise-tolerant audio-visual online person verification using an
  attention-based neural network fusion,''
\newblock in {\em IEEE International Conference on Acoustics, Speech and Signal
  Processing (ICASSP)}, 2019, pp. 3995--3999.

\bibitem{sari2021multi}
Leda Sar{\i}, Kritika Singh, Jiatong Zhou, Lorenzo Torresani, Nayan Singhal,
  and Yatharth Saraf,
\newblock ``A multi-view approach to audio-visual speaker verification,''
\newblock in {\em IEEE International Conference on Acoustics, Speech and Signal
  Processing (ICASSP)}, 2021, pp. 6194--6198.

\bibitem{zhang2016joint}
Kaipeng Zhang, Zhanpeng Zhang, Zhifeng Li, and Yu~Qiao,
\newblock ``Joint face detection and alignment using multitask cascaded
  convolutional networks,''
\newblock {\em IEEE Signal Processing Letters}, vol. 23, no. 10, pp.
  1499--1503, 2016.

\bibitem{kowalski2017deep}
Marek Kowalski, Jacek Naruniec, and Tomasz Trzcinski,
\newblock ``Deep alignment network: A convolutional neural network for robust
  face alignment,''
\newblock in {\em Proceedings of the IEEE conference on computer vision and
  pattern recognition workshops}, 2017, pp. 88--97.

\bibitem{chen2020multi}
Zhengyang Chen, Shuai Wang, and Yanmin Qian,
\newblock ``Multi-modality matters: A performance leap on voxceleb.,''
\newblock in {\em Interspeech}, 2020, pp. 2252--2256.

\bibitem{cortes1995support}
Corinna Cortes and Vladimir Vapnik,
\newblock ``Support-vector networks,''
\newblock {\em Machine learning}, vol. 20, no. 3, pp. 273--297, 1995.

\bibitem{an2021partial}
Xiang An, Xuhan Zhu, Yuan Gao, Yang Xiao, Yongle Zhao, Ziyong Feng, Lan Wu, Bin
  Qin, Ming Zhang, Debing Zhang, et~al.,
\newblock ``Partial fc: Training 10 million identities on a single machine,''
\newblock in {\em Proceedings of the IEEE/CVF International Conference on
  Computer Vision}, 2021, pp. 1445--1449.

\bibitem{liang2018learning}
Zhengfa Liang, Yiliu Feng, Yulan Guo, Hengzhu Liu, Wei Chen, Linbo Qiao,
  Li~Zhou, and Jianfeng Zhang,
\newblock ``Learning for disparity estimation through feature constancy,''
\newblock in {\em Proceedings of the IEEE conference on computer vision and
  pattern recognition}, 2018, pp. 2811--2820.

\bibitem{fan2020cn}
Yue Fan, JW~Kang, LT~Li, KC~Li, HL~Chen, ST~Cheng, PY~Zhang, ZY~Zhou, YQ~Cai,
  and Dong Wang,
\newblock ``Cn-celeb: a challenging chinese speaker recognition dataset,''
\newblock in {\em IEEE International Conference on Acoustics, Speech and Signal
  Processing (ICASSP)}. IEEE, 2020, pp. 7604--7608.

\bibitem{MUSAN}
D.~Snyder, G.~Chen, and D.~Povey,
\newblock ``{MUSAN}: {A} music, speech, and noise corpus,''
\newblock {\em CoRR}, vol. abs/1510.08484, 2015.

\bibitem{RIRS}
Tom Ko, Vijayaditya Peddinti, Daniel Povey, Michael~L. Seltzer, and Sanjeev
  Khudanpur,
\newblock ``A study on data augmentation of reverberant speech for robust
  speech recognition,''
\newblock in {\em IEEE International Conference on Acoustics, Speech and Signal
  Processing (ICASSP)}, 2017, pp. 5220--5224.

\end{thebibliography}
\end{document}